\documentclass[aps,prr,twocolumn,showpacs,floatfix]{revtex4-2}
\usepackage{graphicx}
\usepackage{float,epsfig}
\usepackage{dcolumn}
\usepackage{bm}
\usepackage{graphicx}
\usepackage{amsmath,amssymb,amsthm}
\usepackage{color}
\usepackage[colorlinks=true,linkcolor=blue]{hyperref}

\usepackage{times}
\usepackage{multirow}

{\tiny }

\newcommand{\tchi}{\tilde{\chi}}

\newcommand{\ME}{{\rm EM}}

\begin{document}  

\title{Dimers and discrete breathers in Bose-Einstein condensates in a quasi-periodic potential.  }

\author{
	Vladimir V. Konotop 
}

 \affiliation{
 	Departamento de F\'isica and Centro de F\'isica Te\'orica e Computacional, Faculdade de Ci\^encias, Universidade de Lisboa, Campo Grande 2, Edif\'icio C8, Lisboa 1749-016, Portugal 
}

\begin{abstract} 
	
A quasi-one-dimensional Bose-Einstein condensate loaded into a quasi-periodic potential created by two sub-lattices of comparable amplitudes and incommensurate periods is considered. Although the conventional tight-binding approximation is not applicable in this setting, the description can still be reduced to a discrete model that accounts for the modes below the mobility edge. In the respective discrete lattice, where no linear hopping exists, solutions and their dynamics are governed solely by nonlinear interactions. Families of nonlinear modes, including those with no linear limit, are described with a special focus on dimers, which correspond to breather solutions of the Gross-Pitaevskii equation with a quasi-periodic potential. The breathers are found to be stable for negative scattering lengths. Localization and stable propagation of breathers are also observed for positive scattering lengths at relatively weak and moderate nonlinearities.	
  
\end{abstract}

\maketitle

\section{Introduction}

The simplest model of a quasi-periodic potential in the one-dimensional Schr\"{o}dinger equation is a superposition of two cosine-function sub-lattices with incommensurate periods. This model has received considerable attention in both experimental~\cite{Roati2008,Reeves2014,Luschen2018} and theoretical~\cite{Diener2001,Sakaguchi2006,Modugno09,Adhikari2009,Biddle2010,Yao2019,PraZezKon22} studies of Bose-Einstein condensates (BECs). It is known~\cite{FroSpWi,Surace1990}, that if the depth of such a potential is below a certain value, the respective linear Hamiltonian does not support spatially localized states. Such states appear for higher amplitudes, marking a localization-delocalization transition. However, even in such potentials the localized states are observed only for energies below a certain value, referred to as the mobility edge (ME)~\cite{Mott} while eigenstates of higher energies remain delocalized (several MEs can exist in more complex systems). Both, localization-delocalization transition and ME in one-dimensional BECs loaded in incommensurate cosine-like potentials were studied numerically, see e.g.~\cite{Li2016} and~\cite{Modugno09,Biddle2010,PraZezKon22}, respectively. 

If the incommensurate sub-lattices creating the potential have significantly different amplitudes, then the deepest one creates a potential whose lowest band is well described within the framework of the tight-binding approximation. The shallow lattice introduces an incommensurate modulation, leading to the well-known discrete Aubry-Andr\'e model~\cite{AubryAndre} (AA) (or to one of its generalizations), which is the most commonly used and best-studied single-band discrete quasi-periodic model. 
 
The conventional tight-binding approximation~\cite{tight-binding}, which employs an orthogonal basis of functions localized near the (deep) lattice minima, is no longer applicable when the amplitudes of the sub-lattices are of the same order. Nevertheless, if only the low-energy states, i.e., those belonging to the energies below the ME, are excited, these states can be considered a new basis that spans the wavefunction. Since such a basis is orthogonal, the evolution of an initially localized wavefunction is relatively simple. 

Non-trivial dynamics can be observed only if the coupling of states is introduced either by external factors, such as a weak linear force~\cite{Prates24}, or by inter-atomic interactions. In the latter case, nonlinear hopping may lead to the creation of localized states that are impossible in linear and weakly nonlinear limits, and govern dynamics involving mores that one state. It is noteworthy that, so far, no examples of nonlinear families of the solutions have been reported in quasi-periodic potentials.

In this paper, we consider a Bose-Einstein condensate loaded into a quasi-periodic potential, composed of two optical lattices with incommensurate periods, within the framework of the mean-field approximation. A discrete lattice equation governing low-energy matter waves is derived (Sec.~\ref{sec:lattice}). For a sufficiently general case of spatially localized initial wave packets, the description can be reduced to a few-mode model. We particularly focus on a dimer, corresponding to two excited modes linked by nonlinear hopping, which can be viewed as a two-hump breather. Such a breather has properties very distinct from those known for a dimer of the self-trapping model~\cite{Eilbeck84,Kenkre86,Scott90,Tsironis93}, i.e., a discrete nonlinear Schr\"odinger (DNLS) equation, as well as from a dimer describing atoms in a double-well trap~\cite{Smerzi97,Raghavan1999,GarAbd,TheKev} (Sec.~\ref{sec:dimer}). In addition to families of nonlinear modes bifurcating from the linear localized states, the model considered here, supports families of nonlinear solutions that do not have a linear limit, i.e., existing only if the number of atoms exceeds a critical value. Two-hump breathers exhibit remarkably stable evolution governed by the one-dimensional Gross-Pitaevskii equation (GPE) (Sec.~\ref{sec:dynGPE}).

\section{Lattice equation}
\label{sec:lattice}

We start with a quasi-one-dimensional Gross-Pitaevskii equation (GPE)
\begin{equation}
	\label{GPE}
	i \partial_t\Psi= H\Psi+ g|\Psi|^2 \Psi, \quad H=- (1/2)\partial^2_x+V(x)  
\end{equation}
for the dimensionless order parameter $\Psi(x,t)$ normalized as $\int |\Psi|^2dx=N$ such that $g=+1$ and $g=-1$ correspond to positive and negative scattering lengths of the inter-atomic interactions (the norm $N$ can be viewed as properly normalized number of atoms). The $V(x)$ is a quasi-periodic potential that is deep enough for the existence of a mobility edge (ME) denoted below as $\epsilon_\ME$ (here we consider the case of only one ME). This means that the  eigenstates of the linear stationary Schr\"odinger equation $H\phi_j=\epsilon_j \phi_j $ (numbered by $j=1,...$) with $\epsilon_j \leq \epsilon_\ME$ are localized in space, and thus can be normalized to one $\langle \phi_j,\phi_j\rangle=1$ (hereafter $ \langle f, g\rangle=\int f^*gdx $) while all states with energies above $\epsilon_\ME$ are extended.  

Although a specific choice of the potential remains largely unconstrained, for the sake of definiteness, the detailed analysis below will be presented for the simplest potential created by two optical sublattices 
\begin{align}
	\label{eq:V}
	V(x)=V_1\cos(2x) +V_2\cos(2\beta x+\theta)
\end{align}
with the amplitudes $V_1$ and $V_2$,  the incommensurate relation $\beta$ between the periods (i.e., $\beta$  is an irrational number), and an arbitrary shift $\theta$ used to break the spatial symmetry. This type of potential was exploited in several previous studies~\cite{Diener2001,Modugno09,Yao2019,PraZezKon22}. In the chosen dimensionless units $V_{1,2}$ are measured in the units of the recoil energy, and for typical experimental settings with the scattering length of order of $5\,$nm the norm  $N=1$ corresponds to a few hundreds of atoms. The linear limit $N\to 0$ corresponds to negligible scattering lengths (can be achieved using Feshbach resonance). 

Formally, a quasi-periodic potential (\ref{eq:V}) is defined on the entire real axis, and the respective number of localized modes is infinite. However, the extent of a real-world physical system is finite, although it may greatly exceed the number of periods of each sublattice. By focusing on the dynamics of initial excitations localized far enough from the boundaries in this latter case, one can employ the method of periodic approximants~\cite{Diener2001,Ostlund1983,Modugno09,PraZezKon22,Prates24}. This approach involves replacing the irrational $\beta$ in the truly quasi-periodic potential $V(x)$ with its best rational approximations (BRAs)~\cite{Khinchin} $p_n/q_n$ where $p_n$ and $q_n$ are coprime integers and $n$ is the order of approximation. This gives origin to a periodic potential $V^{(n)}=V_1 \cos(2x)+V_2 \cos{(2(p_n/q_n) x+\theta)}$ with the period $\pi q_n$. Now periodic boundary conditions can be imposed. 

\begin{figure}[h]
 	\includegraphics[width=\columnwidth]{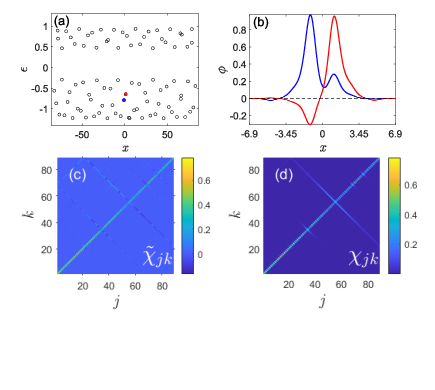}
 	\vspace{-2cm}
 	\caption{(a) Position of the center of mass of all localized modes in the coordinate-energy space. Blue and red asterisks in circles indicate modes $\phi_{32}$ and $\phi_{37}$ used in numerical examples studied in the text and shown in panel (b) with corresponding colors. Two-mode hopping matrices $\tilde{\chi}_{jk}$ and $\chi_{jk}$ for those are shown in panels (c) and (d), respectively (the axes show mode numbers). All results are shown for the potential $V^{(9)}(x)$ (i.e., for $n=9$) with $V_1=1.5$, $V_2=2$, and $\theta=0.13$. For this example there are $M=89$ localized modes (all of them are shown) with the ME  $\epsilon_\ME=\epsilon_{89}\approx 0.9330$; the energy of the first extended mode is $\epsilon_{90}\approx 2.181$. Note the different color scales in panels (c) and (d).}
 	\label{fig:1} 
 \end{figure}

The outlined approximation fairly well describes profile and dynamics of the modes which remain localized inside the interval $I_n=(-\pi q_n/2,\pi q_n/2)$ sufficiently far from its boundaries. Furthermore, in passage from $n$-th to $n+1$-th BRAs, the most important information about solutions of the modes obtained in the $n$-th BRA is preserved in the $n+1$-th BRA (the property termed a memory effect in~\cite{ZezKon22,PraZezKon22}). By the circles in Fig.~\ref{fig:1} (a) the distribution of linear localized states in the coordinate-energy space $(X_j,\epsilon_j)$, where $X_j= \langle \phi_j,x\phi_j\rangle$ is the center of mass (c.m.), is shown. For all numerical illustration in this paper we $n=9$-th  approximant $V^{(9)}(x)$ with $V_1=1.5$, $V_2=2$, and $p_{9}/q_{9}=89/55$ being the $9$-th BRA of the golden ratio $\beta=(\sqrt{5}+1)/2$. The localized modes $\phi_j(x)$ can be chosen real and their localization  is characterized by the inverse participation ratio (IPR) $\langle \phi_j^2,\phi_j^2\rangle\gtrsim 0.1$ [see (\ref{chi}) below].

In Fig.~\ref{fig:1} (a) one observes that the localized modes are distributed nearly homogeneously along the condensate and over the energy axis except one large "gap" (note that although the spectrum is discrete now, the periodic boundary conditions allow one to connect it with the band-gap spectrum of the respective approximant considered on the whole real axis). Upon increase of the order of BRA, the spatial distribution of the modes remains nearly unchanged: new modes appear in the interval $I_{n+1}/I_n$, while density along the energy scales increases with BRA tending to a fractal-like distribution (but without significant changes of the large gaps in the spectrum).  
    
{\color{black} Since localized and extended modes are separated by a gap, when addressing low-energy excitations of the condensate [i.e., modes with $\epsilon\lesssim 1$ in the example shown in Fig.~\ref{fig:1} (a)]  it is sufficient to consider only localized modes. In such a situation, even strong nonlinearity typically does not result in coupling between localized and extended states, as discussed below. Then, }
assuming that the number of localized states is $M$ one can expand $
	\Psi(x,t)=e^{-i \mu t}\sum_{j=0}^{M}a_{j}(t)\phi_j(x),
$
where $\mu$ is the chemical potential and the amplitudes of modes $a_j$ solve the system
\begin{align}
	\label{DNLS}
	i\frac{da_j}{dt}=(\epsilon_j-\mu)a_j+\sum_{j_1, j_2,j_3=1}^M\chi_{jj_1;j_2j_3}a_{j_1}^*a_{j_2}a_{j_3}
\end{align}
the nonlinear coefficients are given by
\begin{align}
	\label{chi}
	\chi_{j_1j_2;j_3j_4}=
	\begin{cases}
		 g\langle \phi_{j_1}\phi_{j_2},\phi_{j_3}\phi_{j_4}\rangle, & \mbox{for $1\leq j_1,j_2,j_3,j_4\leq M$}
		 \\
		 0 & \mbox{otherwise}
	\end{cases} 
\end{align}
and the asterisk stands for complex conjugation. The expansion coefficients are normalized:
$	\sum_{j=1}^M|a_j|^2=N $.

The lattice (\ref{eq:dyn-reduced}) has only nonlinear dispersion lacking the linear one. This imposes constraints on possible quasi-linear states, obtained in the limit $N\to 0$. Indeed, in the limit of negligible nonlinearity, absence of interactions prohibits thermalization of the lattice. This means that for a stationary state in the limit $N\to0$ the chemical potential $\mu$ must approach one of the discrete energies $\epsilon_j$ what corresponds gathering all atoms in a unique state $\phi_j$ (below we discuss this effect in more detail for a dimer). Furthermore, the explicit form of the lattice (\ref{DNLS}) implies nonexistence of the homogeneous distribution of atoms. Indeed, considering $|a_1|=\cdots=|a_M|=a$, where $a=\sqrt{N/M}$, the system (\ref{DNLS}) with $da_j/dt=0$ has $M$ equations but only three two free real parameters $a$ and $\mu$, thus inhibiting solutions for $M>2$ in a general situation.
 	 
{\color{black} Next we simplify Eq.~(\ref{eq:dyn-reduced}). To this end} we give a closer look at the nonlinear hopping integrals. The localized states $\phi_j(x)$ can be chosen real. Thus nonlinear hopping integrals are real too. In Figs.~\ref{fig:1} (c) and (d) we illustrate the matrices $\tilde{\chi}_{jk}$ and $\chi_{jk}$  where   $\chi_{jk}=\chi_{kj}=\chi_{jk;jk}$ and  $\tchi_{jk}=\chi_{jj;jk}$ [for the approximant $V^{(9)}(x)$]. Generally speaking for $\tchi_{jk}\neq\tchi_{kj}$ if $j\neq k$,
and absolute value of at least one of these integrals is larger than $\chi_{jk}$, as it follows from the inequality $\left(\int\phi_j^2\phi_k^2dx\right)^2 \leq \int|\phi_j^{3}\phi_k|dx\int|\phi_j\phi_k^3|dx$. Note, that  $\tchi_{jk}$ is not sign definite. The diagonal elements  $\chi_{jj}=\tchi_{jj}=:\chi_j$ are the IPRs introduced above. One observes that each mode [see the examples in Fig.~\ref{fig:1} (b)] overlaps appreciably with only a very few neighboring modes [upon neglecting small integrals the matrices $\tilde{\chi}_{jk}$ and $\chi_{jk}$ become sparse].  

The nonlinear coefficients involving hopping of three and four different modes, are even much smaller than either $\tilde{\chi}_{jk}$ or $\chi_{jk}$. Therefore we neglect all terms with  $\chi_{jj_1;j_2j_3}$ having three or all four indexes different. In this approximation (\ref{DNLS}) is recast in a form of the reduced lattice model
 \begin{align}
 	\label{eq:dyn-reduced}
 	i\frac{d a_j}{dt} =&(\epsilon_j-\mu)a_j+\chi_j |a_j|^2a_j
 	\nonumber \\	
 	&+
 	\sum_{k\neq j}\left[\tchi_{jk} (2|a_j|^2a_{k}+a_j^2a_{k}^*)
 	\right. \nonumber \\
 	&+	\left. \chi_{jk}(2|a_{k}|^2a_{j}+a_{k}^2a_{j}^*)+\tchi_{kj}|a_{k}|^2a_{k}\right] 
 \end{align}
 We emphasize that the sub-indexes in (\ref{eq:dyn-reduced}) [and in (\ref{DNLS})] indicate the energy level, rather than spatial ordering of the modes, unlike in the conventional AA our DNLS models, where the index stands for the lattice sites.

\section{Nonlinear dimer}
\label{sec:dimer}

The simplest approximate solution of (\ref{eq:dyn-reduced}) is a monomer, when only one, say $j$-th, mode is excited $a_j=\sqrt{N}$ with the chemical potential $\mu_j(N)=\epsilon_j+N\chi_j$.

A dimer corresponds to the choice of only two excited states. The particular relevance of such a reduced model for the lattice (\ref{eq:dyn-reduced}) stems from the results shown in Fig.~\ref{fig:1} (c),(d), where nearly for each arbitrarily chosen state, say $j$-th one, non-negligible  nonlinear coupling can be found only for one (or a very few) other sites, say $k$-state. Thus, either $|\tchi_{jk}|$ or $|\tchi_{kj}|$ coefficients are much larger than other nonlinear hopping integrals. Having chosen two nonlinearly coupled states $j$ and $k$ while neglecting all other, the system (\ref{eq:dyn-reduced}) is reduced to a dimer. Like other dimer models studied previously~\cite{Smerzi97,Raghavan1999,GarAbd,PraZezKon22}, the dimer considered here is conveniently described by the ansatz 
\begin{align}
	\label{ansatz}
a_j=\sqrt{\frac{N(1+z)}{2}}e^{i(\theta-\varphi)/2}, \,\, a_k=\sqrt{\frac{N(1-z)}{2}}e^{i(\theta+\varphi)/2}
\end{align}
where $z(\tau)$ is the population imbalance, $2\varphi(\tau)$ is the phase mismatch, and $\theta(\tau)$ is a global rotating phase, as well as, the re-scaled time $\tau=N\chi_{jk}t$. Setting the chemical potential as
\begin{align}
	\mu=\mu_j(N/2)+ \mu_k(N/2)+\chi_{jk}
\end{align}
one verifies that the evolution of $z(\tau)$ and $\varphi(\tau)$ does not depend on $\theta(\tau)$ and is governed by  Hamilton's equations  
\begin{align}
	\label{HamZ}
	\frac{dz}{d\tau}=(1-z^2)\sin(2\varphi)+\sqrt{1-z^2}(\eta_-z+\eta_+)\sin\varphi
	\\
	\label{HamPhi}
	\frac{d\phi}{d\tau}=\nu+\xi_-	-(1-\xi_+)z-2 z\cos^2\varphi 
	\nonumber \\
 -\frac{2\eta_-z^2+\eta_+z-\eta_- }{\sqrt{1-z^2}}\cos\varphi
\end{align}
with the Hamiltonian given by
\begin{align}
	\label{Hamilt}
	H=(1-z^2)\cos^2\varphi+\sqrt{1-z^2}(\eta_-z+\eta_+)\cos\varphi
	\nonumber \\
	-\frac{1-\xi_+}{2} z^2+(\nu+\xi_-) z
\end{align}
and parameters defined by
\begin{align*}
\nu= \frac{\epsilon_j- \epsilon_k}{\chi_{jk} N},   \quad \xi_\pm=\frac{\chi_j\pm\chi_k}{2\chi_{jk}},
	\quad \eta_\pm=\frac{\tchi_{jk}\pm \tchi_{kj}}{\chi_{jk}}.	 
\end{align*}  
The dependence of the global phase on time is obtained  from the compatibility of (\ref{HamZ}) and (\ref{HamPhi}) with (\ref{eq:dyn-reduced}) written for two modes, and has the form 
\begin{align}
	 \label{eq:theta}
	 \theta=
	 \int_0^\tau\left(   \frac{\eta_+z^2-\eta_-z-2\eta_+}{\sqrt{1-z^2}}\cos\varphi- 2\cos^2\varphi -\xi_-z \right) d\tau' 
\end{align}
[it is set $\theta(0)=0$]. Thus, unlike in most previous dimer models, the dimer described here features a global phase that varies over time. Since this phase does not affect the populations and superfluid currents, it will not be considered in detail below.
 

One can prove that fixed points of system (\ref{HamZ}), (\ref{HamPhi}) exist only for $\varphi=0,\,\pi$ (mod $2\pi$), what readily gives algebraic equation for determining population imbalances of the stationary solutions: 
\begin{align}
	\label{eq:z}
	\nu=(1-\xi_+)z-\chi_- +\sigma\frac{ \eta_+z^2-\eta_-z-2\eta_+}{\sqrt{1-z^2}}.
\end{align}
Here $\sigma=+1$ and $-1$ stand for in-phase ({\it alias} unstaggered), $\varphi=0$, and out-of-phase (staggered), $\varphi=\pi$, superposition of the states $j$ and $k$, respectively.  

Since the number of atoms $N$ enters (\ref{eq:z}) through the parameter $\nu$, Eq.~(\ref{eq:z}) yields dependence $N(z)$. Subsequently, having found $\mu (z)$, one obtains a parametric form of the function $N(\mu)$ describing the families of solutions. Analytical expression of $N(\mu)$ is simple but cumbersome. {\color{black} Therefore we illustrate the dependence  $N(\mu)$ graphically} in Figs.~\ref{fig:1} (a) and (c), for two modes $\phi_{j}=\phi_{32}$ and $\phi_{k}=\phi_{37}$ depicted in Fig.~\ref{fig:1} (b) (this pair of nonlinearly coupled modes was chosen arbitrarily: note that non of them is the ground state).
\begin{figure}[h]
	\includegraphics[width=0.45\columnwidth]{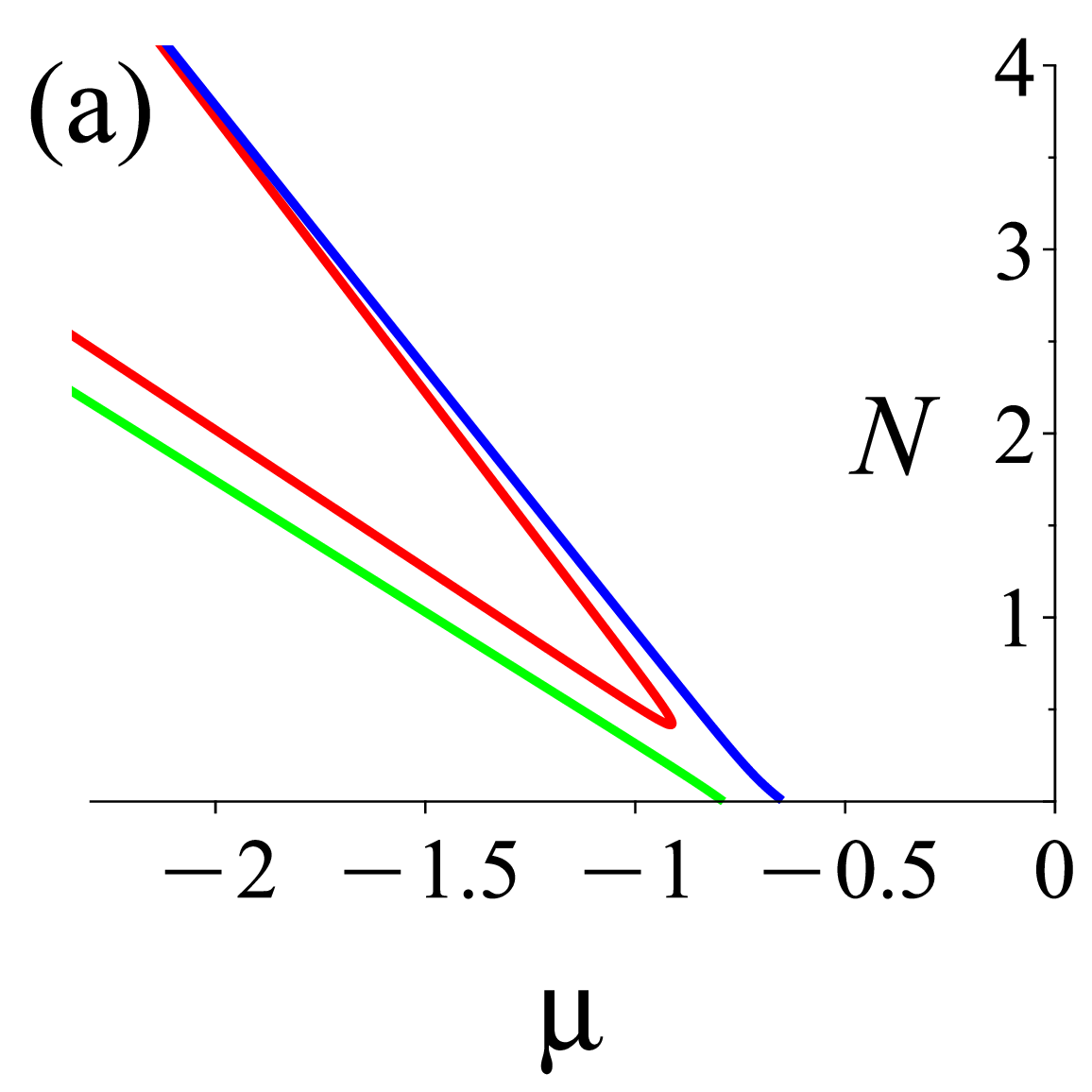}%
	\includegraphics[width=0.45\columnwidth]{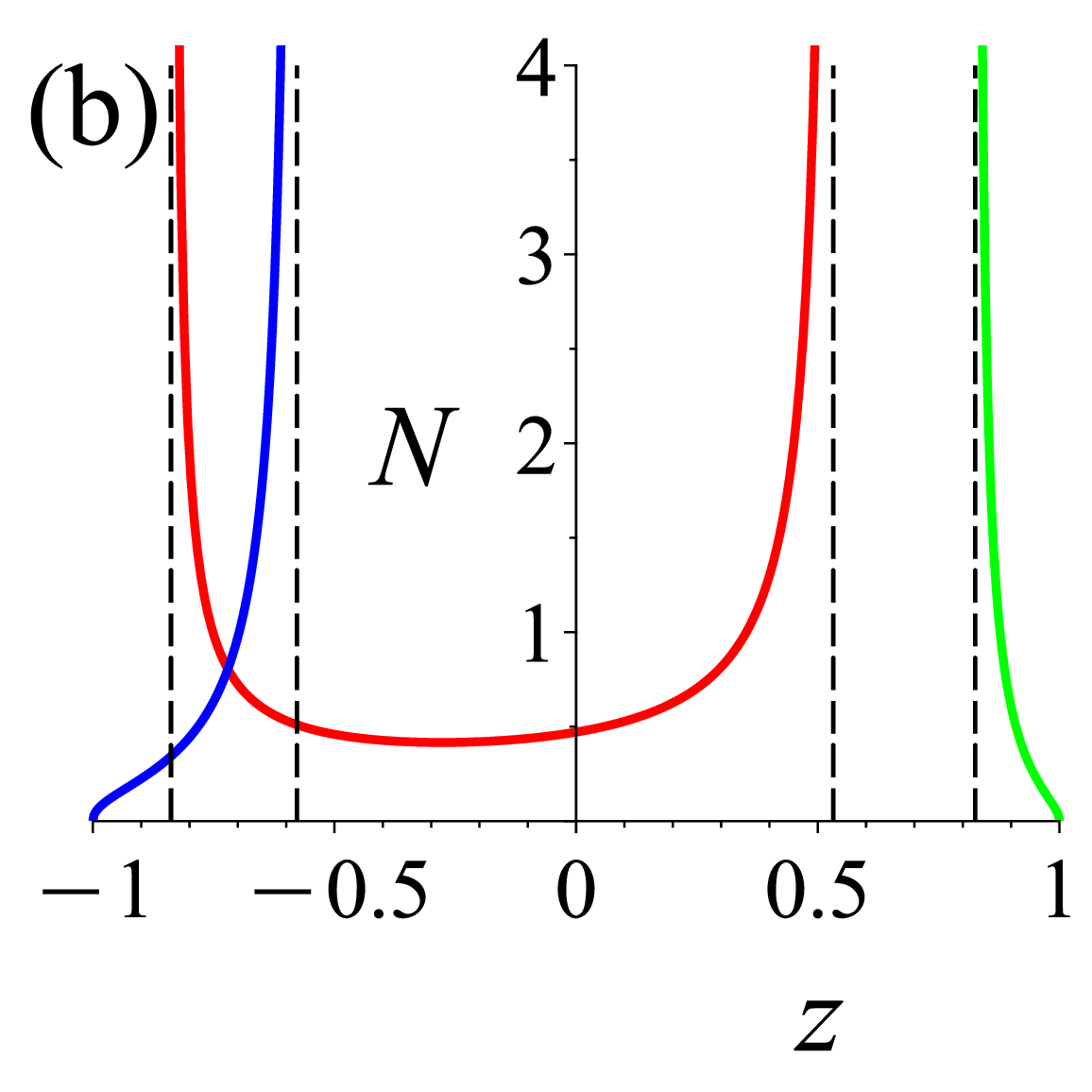}%
	\\
	\includegraphics[width=0.45\columnwidth]{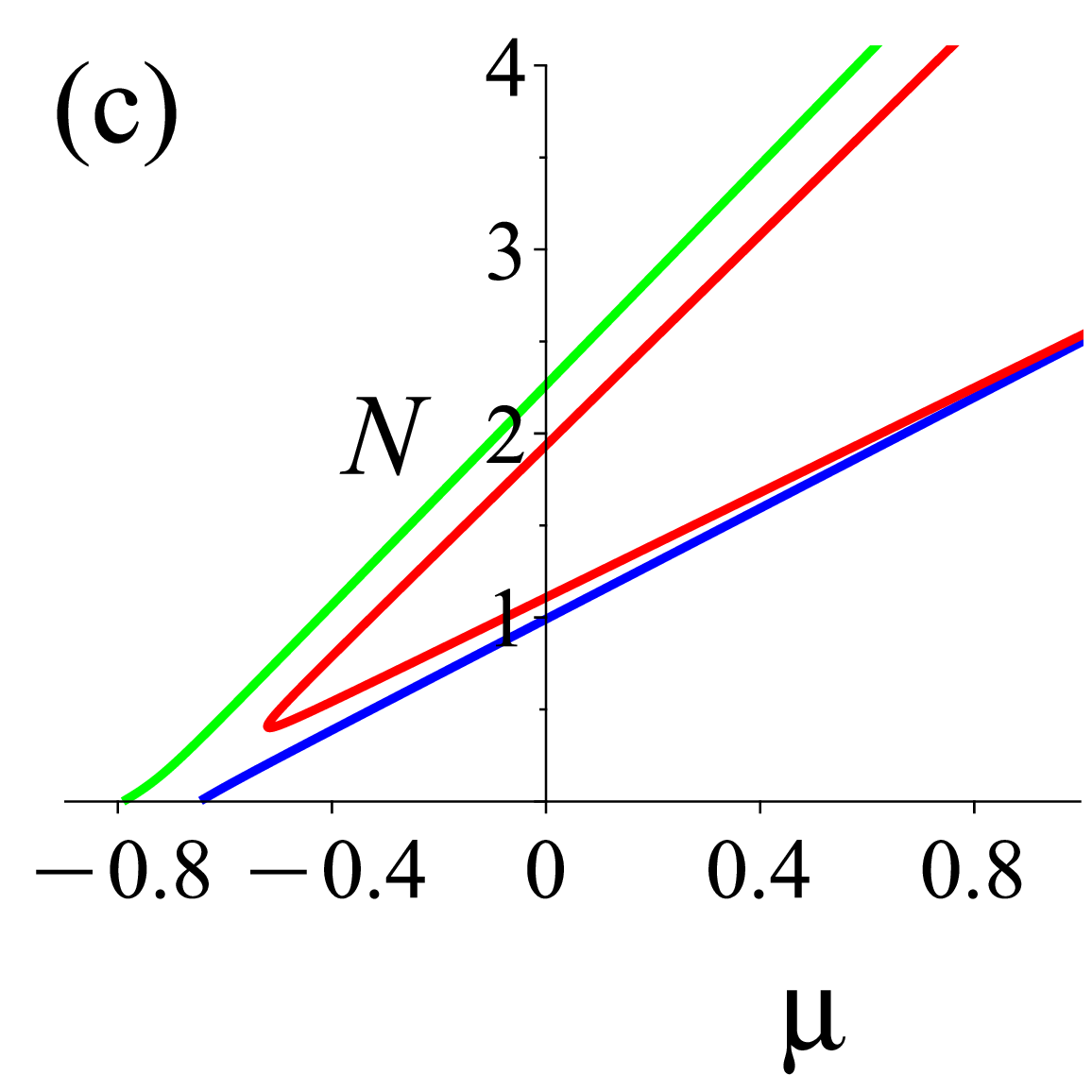}%
	\includegraphics[width=0.45\columnwidth]{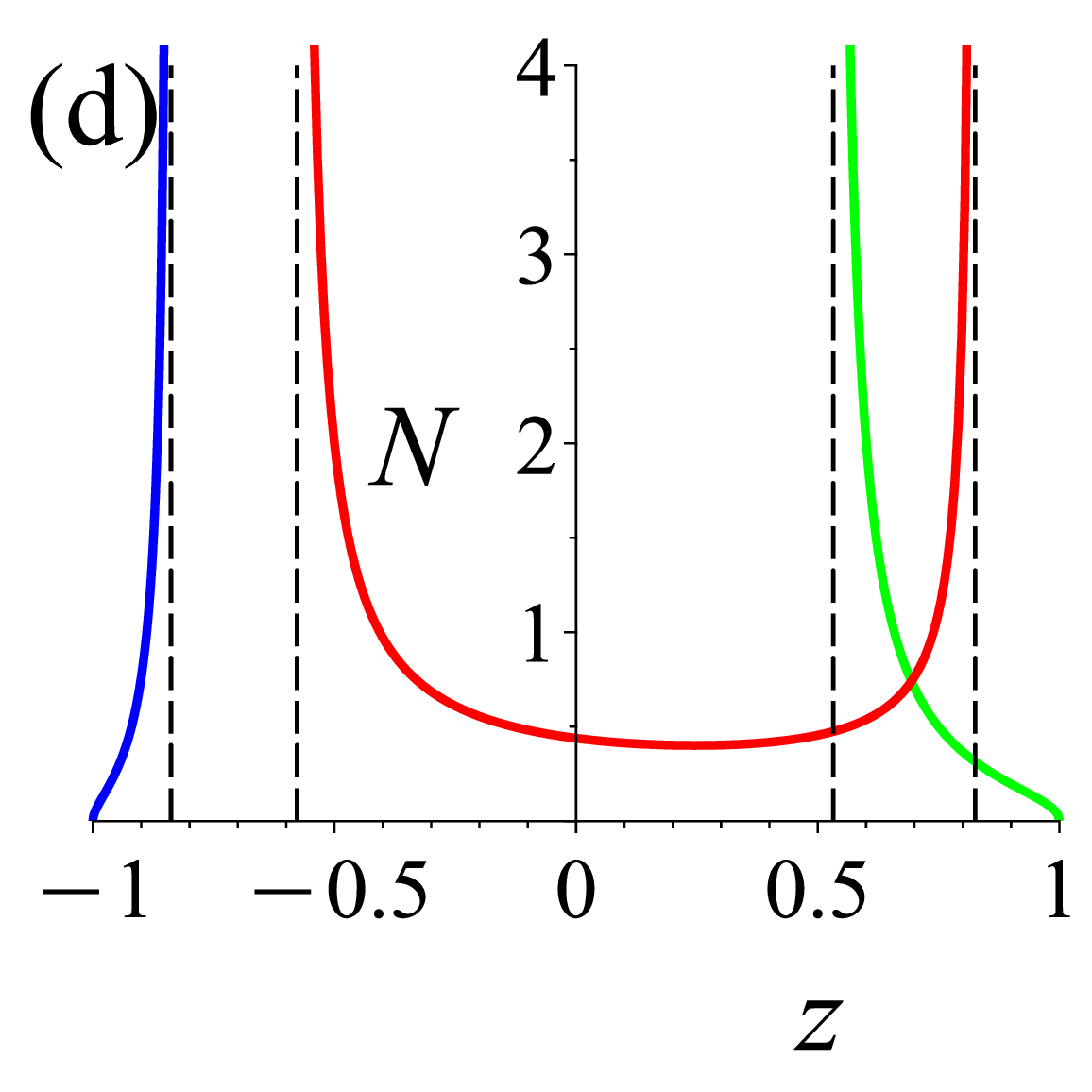}
	\caption{Families of the nonlinear modes in (a), (c) and dependence of the number of atoms $N$ on the imbalance $z$ in (b), (d) are shown for negative (upper panels) and positive (lower panels) scattering lengths. The green (unstaggered) and blue (staggered) families bifurcate from the linear states  $\phi_{32}$ and $\phi_{37}$ respectively. The mixed state having no linear limit is shown by the red line. Vertical dashed lines in (b) and (d) show population imbalances at which $N$ diverges. The parameters of the potential are the same as in  Fig.~\ref{fig:1}. }
	\label{fig:2} 
\end{figure}

Figures~\ref{fig:2} (a) and (c) show the families of solutions for attractive and repulsive interactions, respectively. In both cases there are two families bifurcating from the single linear states $\phi_{32}$ (green line) with $\epsilon_{32}\approx -0.790$ and $\phi_{37}$ (blue line) with $\epsilon_{37}\approx -0.647$. No superposition of such states is possible for relatively small number of particles. Physically this is a consequence of the absence of linear hopping. Mathematically, this follows from the limit $N\to 0$ corresponding to $|\nu|\to\infty$ in (\ref{eq:z}), which is possible only if $|z|\to 1$. In the linear limit it is straightforward to compute the bifurcation angles of the families $N\approx \chi \mu$. These slopes in Figs.~\ref{fig:2} (a) and (c) for attractive ($g=-1$) and repulsive ($g=1$) nonlinearities are given by $\chi_{32}\approx g\cdot 0.589$ and $\chi_{37}\approx g\cdot 0.567$. Note, however, that the effect of both modes are accounted at any $N>0$: the blue and green nonlinear families are staggered and unstaggered modes for negative and positive scattering length, respectively.  When the number of atoms increases, the mentioned families deviate from each other. The results in Fig.~\ref{fig:2} (c) resemble the known families of solutions of a dimer obtained in Ref.~\cite{Eilbeck84} with the difference that now the staggered mode has higher energy than unstaggered one. A peculiarity of the dimer (\ref{HamZ}), (\ref{HamPhi})  appears when the number of atoms is larger than a certain threshold value $N_{\rm th}$ Eq.~(\ref{eq:z}). Then two additional real roots of (\ref{eq:z}) corresponding to two new families emerging through a saddle-node bifurcation. For the parameters used in Fig.~\ref{fig:2} (a) and (c)  $N_{\rm th}\approx 0.416$ and $N_{\rm th}\approx 0.401$, respectively. The modes of the upper family are unstaggered in the case of attractive interactions [Fig.~\ref{fig:2} (a)] and staggered in the case of repulsive interactions [Fig.~\ref{fig:2} (c)] i.e., the upper family modes have a type opposite to the modes bifurcating from the linear limit with which the intersect on the diagrams $(z,N)$ shown in (b) and (d)].

The total number of atoms $N$ defines possible imbalances of populations of the modes as shown in Figs.~\ref{fig:2} (b) and (d) for attractive and repulsive interactions, respectively. Now it is possible to obtain a mode with equally populate states (corresponding to $z=0$), which belongs the higher families without linear limit. We also observe that there exist intervals of imbalances inhibited for any $N$. 

\begin{figure}[h]
	\includegraphics[width=\columnwidth]{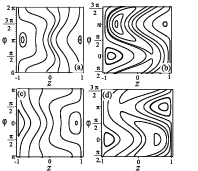}%
	\caption{Phase portraits for attractive ($g=-1$) in (a), (b) and repulsive ($g=1$) in (c), (d) nonlinearities, shown for $N=0.3$ in (a), (c) and $N=2$ in (b), (d),i.e., below and above the threshold nonlinearities (see the text). The parameters of the discrete lattice are obtained for the modes shown in Fig.~\ref{fig:1} (b). }
	\label{fig:3} 
\end{figure}

To describe the dynamics of the dimer in Fig.~\ref{fig:3} we present the phase portraits for attractive (upper panels) and repulsive (lower panels) interactions. The numbers of atoms is below the bifurcation threshold $N=0.3 <N_{\rm th}$  in the left column and above bifurcation threshold $N=2 >N_{\rm th}$ in the right column. Below the threshold value $N_{\rm th}$, as one could expect from the dependencies $N(z)$ in Fig.~\ref{fig:2} (b) and (d), the dynamics is characterized by relatively weak changes in the population imbalance in both cases: weak oscillations around stable points and rotation with the evolving phase $\varphi(t)$ [see Fig.~\ref{fig:3} (a) and (c)]. However, the population of states may not fully depict the atomic density distribution in physical space due to interference effects between states (as seen in Fig.~\ref{fig:1} (b). This interference effect is highlighted in the example presented in Fig.~\ref{fig:4} (a$_1$), where atoms undergo significant transfer between the two spatial locations. Comparing panels (a) and (c) in Fig.~\ref{fig:3}, we observe qualitatively similar dynamics for both types of nonlinearities.

Figures~\ref{fig:3} (b) and (d) show phase portraits of the dimer for $N=2>N_{\rm th}$. In each panel there are three centers [in (b) they are approximately at $(\varphi,z)=(0,-0.80)$, $(\pi, -0.64)$, and $(\pi, 0.85)$] and one hyperbolic fixed point [at $(\varphi,z)=(0, 0.45)$ in panel (b)].  Consequently, both families emerging from the linear limit [represented by the blue and green lines in  Fig.~\ref{fig:2} (b)] are stable. The families without a linear limit are stable if $dN/dz<0$ for attractive interactions and $dN/dz>0$ for repulsive interactions, while they are unstable otherwise. It is noted that the amplitude of variation of $z$ is significantly larger for strong nonlinearities compared to weak nonlinearities.

\section{Two-hump breathers}
\label{sec:dynGPE}

Apart from the neglected hopping between three and four states, the discrete model (\ref{eq:dyn-reduced}) fails to account for the effect of spatial dispersion, which is different for positive and negative scattering lengths. Specifically, the stability of localized states in the GPE (\ref{GPE}) generally differs from the dynamical stability discussed in the preceding section. Moreover, as previously mentioned, the interference of nonlinearly interacting modes can significantly impact the spatial distribution of atomic density in real space due to the finite spatial extension of these modes. Consequently, a more complete characterization of real-space evolution involves studying superfluid current densities alongside the spatial distribution: 
\begin{align}
	\label{current}
	J=\frac{1}{2i}\left(\Psi^*\Psi_x-\Psi\Psi_x^*\right).
\end{align}
Additionally, the dynamics reported in Fig.~\ref{fig:3} does not explicitly show dependence on time, while the number of atoms determines both the system parameters and the time scale $\tau=\chi_{jk} Nt$.  
  
To address the above issues, we consider direct evolution of solutions obtained numerically from the GPE (\ref{GPE}) with the initial conditions constructed based on the dimer solution: 
\begin{align}
	\label{init}
	\Psi (x,0)=\sqrt{\frac{N(1+z_0)}{2}}\psi_j(x)+e^{i\varphi_0}\sqrt{\frac{N(1-z_0)}{2}}\psi_k(x).
\end{align}
Here $z_0$ and $\varphi_0$ are the initial imbalance of atomic population and phase mismatch of the states, respectively. In the particular examples below, $j=32$ and $k=37$. In all numerical results reported below, a noise perturbation of approximately 3\% of the input amplitude was added to the initial condition (\ref{init}). 
    
\begin{figure}[h]
	\includegraphics[width=\columnwidth]{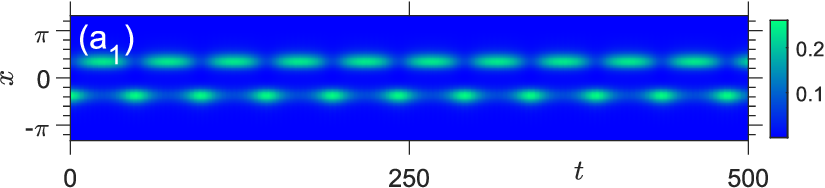} 
	\\ 
	\includegraphics[width=\columnwidth]{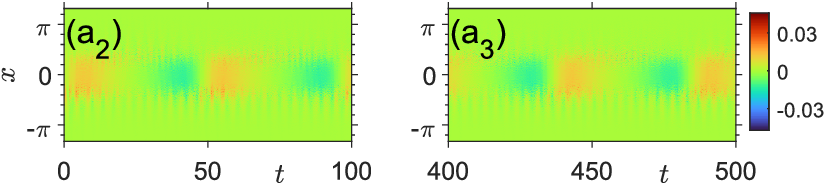}%
	\\
	\includegraphics[width=\columnwidth]{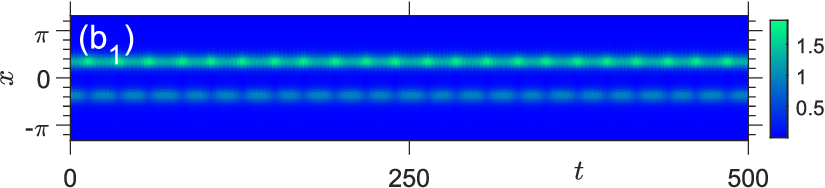} 
	\\ 
	\includegraphics[width=\columnwidth]{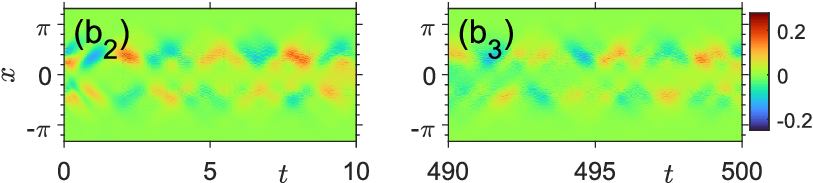}%
 
	\caption{{\color{black} Evolution of the two-mode initial condition (\ref{init}) for a negative scattering length. In panels (a) $z_0=0$, $\varphi_0=\pi$, $N=0.3$; (a$_1$) shows the density over long-time ($T_{\rm fin}=500$) evolution, (a$_2$) and (a$_3$) show the current density over the two initial and two final (computed) periods. Panels (b) show the evolution of the density (b$_1$) and the current density over the two initial (b$_2$) and two final (b$_3$) periods for the initial condition (\ref{init}) with $z_0=0.451$, $\varphi_0=0$ and $N=2$.} The system parameters are the same as in Fig.~\ref{fig:1}.}
	\label{fig:4} 
\end{figure}

Starting with the attractive nonlinearity, it was verified that the stationary states, i.e., dynamically stable families in Fig.~\ref{fig:2} (a) [and centers in Fig.~\ref{fig:3} (a)], remain stable in the direct evolution governed by the GPE (not show here).   Figure~\ref{fig:4} (a) demonstrates the oscillatory behavior of the rotating-phase solution with a number of atoms below the threshold $N_{\rm th}$. Although at $t=0$ the modes $\phi_{32}$ and $\phi_{37}$ [$z(0)=0$] are equally populated, there is a significant imbalance of atoms near the c.m. of the modes [depicted in Fig.~\ref{fig:1} (a)]. The higher energy mode with $k=37$ is, on average, more populated than the lower energy mode with $j=32$, what is a manifestation of the effect of interference.  

While the oscillatory dynamics in Fig.~\ref{fig:4} (a) resembles the one reported in~\cite{PraZezKon22} for a boson-Josephson junction, there is a substantial difference. The boson-Josephson oscillations occur due to the superposition of two linear states, whereas oscillations reported here are enabled by the nonlinearity and vanish at $N\to 0$. This decrease, and eventually vanishing of oscillations with decreasing $N$ can be observed in comparison of the oscillation periods in panels (a) and (b) of Fig.~\ref{fig:4}: $T_{\rm osc}\approx 50$  for $N=0.3$ and $T_{\rm osc}\approx 5$ for $N=2$, respectively. The nonlinear nature of the respective oscillatory solutions suggest their interpretation as two-hump {\em breathers}.   

In the evolution shown in Fig.~\ref{fig:4} (b), where $N=2>N_{\rm th}\approx 0.416$ the initial condition corresponds to the hyperbolic fixed point in Fig.~\ref{fig:3} (b). The instability of  dynamical system (\ref{HamZ}), (\ref{HamPhi}) leads to excitation of a stable two-hump breather. The higher energy mode remains more populated at any time. 

The evolution of the breathers is accompanied by alternating superfluid currents, as shown in panels in panels (a$_{2,3}$) and (b$_{2,3}$) for the initial and advanced time intervals. In the former case, due to atomic exchange, the current density is maximal between the two spatial locations of the c.m. of the modes. In the last case, where the breather corresponds to an unstaggered mode of the dimer, the maxima of current densities remain localized near the c.m. of the modes, which are oppositely directed. In both cases, the directions of the current densities periodically vary over time. 

\begin{figure}[h]
	\includegraphics[width=\columnwidth]{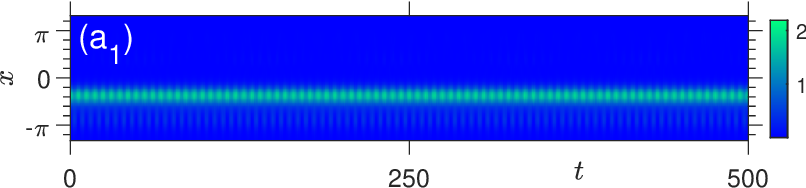} 
	\\
	\includegraphics[width=\columnwidth]{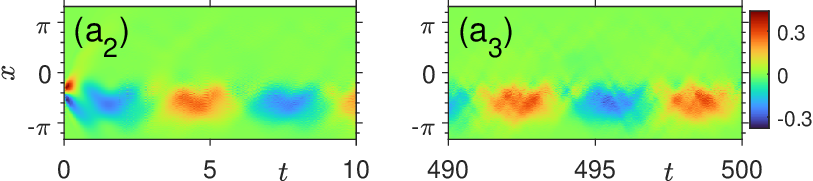}%
	\\
	\includegraphics[width=\columnwidth]{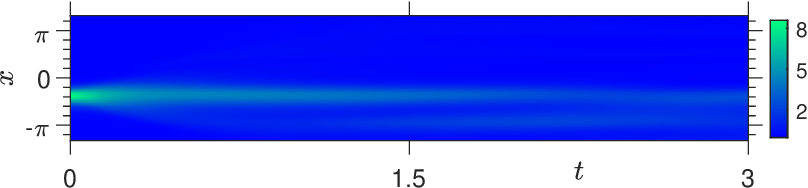} 
\caption{Evolution of the mode corresponding to fixed point $z(0)\approx 0.7886 $, $\varphi(0)=\pi$   and $N=2$ [see Fig~\ref{fig:2} (d)] is shown in panels (a). Panels with indices $1$, $2$ and $3$ show the long-time ($T=500$) evolution of the density $\rho(x,t)$, currents at the initial and final stages of evolution, respectively.
(b) Evolution of the same initial wavepacket but for $N=8$. In both cases $N=2$ and the system parameters the same as in Fig.~\ref{fig:1}. }
	\label{fig:5} 
\end{figure}

Peculiarities of the evolution of breathers in the case of repulsive inter-atomic interactions are shown in Fig.~\ref{fig:5}. Panels (a) illustrate the evolution of a staggered breather mode. Almost all atoms are concentrated in the lower-energy mode $\phi_{32}$ without manifesting transfer to the second mode. This is confirmed by both density in Fig.~\ref{fig:5} (a$_1$) and alternating-current density in Fig.~\ref{fig:5} (a$_{2,3}$), thus offering a quite different evolution pattern as compared with those shown in Fig.~\ref{fig:4}. In spite of repulsive interactions and moderate number of particles ($N=2$) there is no dispersive spreading of the breather, that would be observable over large time intervals.  

Meantime, for the chosen parameters of the system and of the pair of modes, the dispersion occurring due to the repulsive nonlinearity becomes visible at shorter evolution times when higher nonlinearities are present. In Fig.~\ref{fig:5} (b) the fast dispersive decay of the breather is shown for $N=8$.
 
\section{Discussion and conclusion}

We have shown that the description of matter waves in one-dimensional quasi-periodic potentials created by optical lattices of comparable amplitudes can be reduced to a discrete model for low-energy initial states. This lattice model features nonlinear dispersion without a linear counterpart. Consequently, in the limit of a relatively small density of atoms, its stationary solutions are represented only by families bifurcating from the single linear modes, while high nonlinearities lead to the emergence of upper families of nonlinear modes. Interestingly, in this sense, the lattice model can be viewed as opposite to the known self-trapping (or discrete nonlinear Schr\"odinger type) models, where the anti-continuum limit~\cite{MacKay1994}, i.e., the limit of completely decoupled modes, occurs at formally infinite nonlinearities. In our model, such an anti-continuum limit occurs at zero nonlinearities. On the other hand, a similarity between both types of models is observed in the property of a growing number of possible nonlinear localized states with increasing nonlinearity.

From the experimental point of view, in the limit of very large (infinite) nonlinearities  the lattice model (\ref{DNLS}) [and respectively (\ref{eq:dyn-reduced})] fails to describe the full dynamics of the BEC in a quasi-periodic potential, because localized and extended states become coupled. However, this occurs at excessively high values of $N$, making the range of applicability of the model large enough. Indeed, if initially only localized states are excited, the effect of extended ones on the dynamic is determined by the hopping between the (normalized) states below and above ME. For a system with spatial extension $L$, amplitudes of extended states are $\sim 1/\sqrt{L}$ (due to the normalization factor) which approximately determines the amplitude of their hopping with localized states. Such nonlinear hopping can be neglected compared with the nonlinear hopping between localized states if $1/\sqrt{L}\ll 1 $.  In the dimensional units for the 9-th BRA of the golden ratio used here  $1/\sqrt{L}=1/\sqrt{\pi q_n}\approx 0.076$, and rapidly decreases with the order of BRA. This explains, in particular, remarkable stability of the breathers observed in the full-scale dynamical simulations even for finite magnitudes of both attractive and repulsing nonlinearities.

The approach adopted in this paper is based on periodic approximants. Validity of the approach of periodic approximants, adopted in this paper, can be justified by the so-called memory effect~\cite{ZezKon22,PraZezKon22} which, loosely speaking, means that the $n+1$-th approximant introduces only small corrections to the results obtained for the $n$-th approximant within a single spatial period. These corrections become negligible with increasing order of the BRA. Meanwhile, the setting using periodic boundary conditions can be viewed, alternatively, as a quasi-one-dimensional description of a BEC in a toroidal trap (experimentally such trap are created routinely, see e.g.~\cite{Henderson2009,Amico2022}) in the presence of an additional periodic potential (similar to one considered in~\cite{Saito2004}). 

Finally, the theory developed here for a quasi-one-dimensional BEC, can be extended to two- and three-dimensional nonlinear systems which are now experimentally available. Examples include BECs in twisted lattices~\cite{Meng2023} and in quasicrystals~\cite{Hou2018,Yu2023}, as well as light propagation in nonlinear photorefractive moir\'e lattices~\cite{Fu2020} when, strictly speaking, the conventional tight-binding approximation is not applicable any more.

 \begin{acknowledgments}
 The work was supported by the Portuguese Foundation for Science and Technology (FCT) under Contracts UIDB/00618/2020 (DOI: 10.54499/UIDB/00618/2020) and  PTDC/FIS-OUT/3882/2020 (DOI: 10.54499/PTDC/FIS-OUT/3882/2020).  
 \end{acknowledgments}

\end{document}